\documentstyle[aps,prl,multicol,epsfig]{revtex}

\newcommand{\be}{\begin{eqnarray}}
\newcommand{\ee}{\end{eqnarray}}

\draft
\tighten

\begin{document}
\title{Vortices in Trapped Superfluid Fermi-gases}
\author{M.\ Rodriguez$^1$, G.-S.\ Paraoanu$^2$ and P.\ T\"orm\"a$^1$}
\address{$^1$Laboratory of Computational Engineering, P.O.Box 9400, FIN-02015
Helsinki University of Technology, Finland  \\
$^2$ Department of Physics, Loomis Laboratory, 1110 W.\ Green Street,
University of Illinois at Urbana-Champaign, Urbana IL61801, USA}
\maketitle

\begin{abstract} 
We consider a superfluid of trapped fermionic atoms and study
the single vortex solution in the Ginzburg-Landau regime.
We define simple analytical estimates for the main characteristics
of the system, such as the vortex core size, temperature regimes for
the existence of a vortex, and the effects of rotation and interactions 
with normal fermions. The parameter
dependence of the vortex core size (healing length) is found to 
be essentially different from that of the healing length in metallic 
superconductors or in trapped atomic BEC in the Thomas-Fermi limit. 
This is an indication of the importance of the confining geometry 
for the properties of fermionic superfluids. 
\end{abstract}


\begin{multicols}{2}[]

Experiments on cooling of trapped gases of Fermionic atoms
  \cite{Debbie,Salomon,Thomas} are at the level where the achievement of the
  predicted BCS-transition \cite{Stoof,Holland} can be anticipated.
  Cooper-paired trapped atoms would allow to study and test the
  BCS-theory in a controlled manner -- for
  instance the classic problem of the BCS-BEC crossover when the
  interparticle attraction varies \cite{bcsbec} could be studied using the
  possibility to tune the interatomic scattering length.  
  Several methods for observing the existence of a gap in the
  excitation spectrum of the superfluid Fermi-gas have been 
  proposed \cite{all,ours}.
  
  Vortices are a macroscopic signature of the superfluidity, and the
  vortex core size reflects the typical coherence lengths of the
  system. Vortices have played a major role in experimental and
  theoretical studies of superfluid Helium\cite{He} 
  and lately, of condensates of Bosonic atoms \cite{boson}.
  We consider vortex solutions for superfluid trapped
  Fermi-gases. Although a full description of the system would require
  a careful investigation of the Bogoliubov-deGennes equations
  \cite{BdG,Ivan}, we have chosen as a first attempt to
  characterize the system using the Ginzburg-Landau equation in a
  trapped geometry \cite{Baranov}.  This allows us to define intuitive
  estimates for the vortex core size and for the temperature regimes
  where a vortex solution exists. Furthermore, the effects of rotation
  as well as of interactions with normal fermions (position-dependent
  Hartree-fields) can be incorporated into the formalism. We estimate
  their effects on the critical temperature and the vortex solution,
  and develope a
  self-consistent method for treating the effect of the Hartree fields.
  We also discuss how our results can be used in
  estimating the possibilities for observing a vortex. 

The Ginzburg-Landau (GL) equation for  atoms in two 
different hyperfine states 
confined in a symmetric trap of frequency $\Omega$ is given by \cite{Baranov}
\be
\left[-\frac{7\zeta(3)}{48\pi^2}{\left(\frac{\hbar\Omega}{k_BT}\right)}^2
{\nabla}^{2}+
\frac{1+2\lambda}{2\lambda}
R^{2}-ln\frac{T_c^{(0)}}{T}\right]\Delta(R) \nonumber 
\\ 
+ \frac{7\zeta(3)}{8\pi^2}\frac{{|\Delta(R)|}^2}{({k_BT})^2}\Delta(R)=0 ,
\ee
where $\lambda=2p_F^{(0)}|a|/(\hbar \pi)$ is the interaction parameter
($< 1$ for dilute systems), $T_c^{(0)}$ is the critical 
temperature for the corresponding homogeneous system,
 $\Delta(R={\bf r}/R_{TF})$ is the spatially dependent order parameter 
and $R_{TF}$ is the cloud size $R_{TF} =(2E_F/(m\Omega^2))^
{1/2}$ given by the Thomas-Fermi (TF) approximation \cite{Tf}.

We impose a singly quantized vortex solution for the order
 parameter in cylindrical coordinates in a general anharmonic trap potential:
$\Delta(R)=\Delta(\rho,z,\theta)=e^{i\theta}\Delta(\rho,z)$ where 
$\rho \equiv \frac{\rho}{R_{TF}}$, $z \equiv \frac{z}{R_{TF}}$ and the 
equation for the real order parameter becomes (in $\hbar = 1$ units)
\be
[-\frac{1}{2m^*}(\frac{\partial^2}{\partial
 \rho^2}+\frac{1}{\rho}\frac{\partial}{ \partial \rho}
+\frac{\partial^2}{\partial z^2}-\frac{1}{\rho^2}) 
+ \frac{m^* {\omega^*}^2}{2}(\rho^2 + Az^2)  \nonumber \\
 - E_{\mu}(T)]\Delta(\rho,z) + C(T) 
{|\Delta(\rho,z)|}^2\Delta(\rho,z)  = 0  ,
 \label{tres}
\ee
where 
\be
m^* &=&  \frac{1}{2}\frac{48\pi^2}{7\zeta(3)}{\left(\frac{k_B T_c}{\Omega_{ho}}
\right)}^2 \left(\frac{T}{T_c}\right)^2\\
\omega^* &=& \sqrt{\frac{1+2\lambda}{\lambda m^*}} \\ \label{freq}
C &=& \frac{7\zeta(3)}{8\pi^2}\left(\frac{T_c}{T}\right)^2 \\
E_{\mu}(T) &=& ln\frac{T_c^{(0)}}{T}=ln \frac{T_c}{T} +\frac{3}{2}\tilde{\omega},
\ee
and $\frac{3}{2}\tilde{\omega}=ln\frac{T_c^{(0)}}{T_c}$.
We use the scaling $\Delta(R) \equiv \Delta (R) /(k_B T_c)$.
The asymmetry of the trapping potential is described by 
$A=\frac{\Omega_{\bot}^2}{\Omega_z^2}$,
and $\Omega_{ho}=(\Omega_{\bot}^2\Omega_z)^{1/3}$.
Thus the GL equation
becomes analogous to the Gross-Pitaevskii (GP) equation for bosonic atoms \cite{boson}
 for particles of an 
effective mass $m^*$ in a harmonic confining potential with an
effective frequency $\omega^*$ and $\omega^*_z=\omega^* \sqrt{A}$.
The nonlinear term plays the role of repulsive 
interparticle interaction, and $E_\mu(T)$ is the ``chemical
potential'' of the system.
Strong superfluidity, compared to trap energy scales, 
means large effective mass and small effective trapping frequency. 

We consider a two-dimensional geometry where the single vortex is uniform 
along its axis ($z$-axis). 
Eq.(\ref{tres}) then reads
\be
[-\frac{1}{2m^*}(\frac{\partial^2}{\partial
\rho^2}+\frac{1}{\rho}\frac{\partial}{ \partial \rho}
-\frac{1}{\rho^2}) + \frac{m^* (\omega^*)^2}{2} \rho^2 \nonumber 
\\ - E_\mu(T)]\Delta(\rho)
+ C(T) {|\Delta(\rho)|}^2\Delta(\rho)=0 \label{2dim} ,
\ee
and now $E_\mu(T) \equiv \ln (T_c/T) + \tilde{\omega}$.
We have used the steepest descent method described e.g.\ in
\cite{method} using imaginary time propagation to search for the ground state
 solution of 
Eq.(\ref{2dim}). We applied the Crank-Nicholson finite differencing method used to 
solve the GP equation, but without normalization after each step as the size of 
$\Delta$ is determined by the nonlinear term. 

{\it Vortex core sizes}
The size of the vortex core reflects the healing length of a superfluid
because within this distance, the order parameter ``heals'' from zero
up to its bulk value. The first guess to estimate the healing length
in our case would be to
equate the ``kinetic energy'' term in Eq.(\ref{2dim}), 
$\sim 1/(2m^*\xi^2)$ to the ``interaction energy", as 
was done in case of trapped atomic BEC 
(GP equation) \cite{Pethick}. This yields ($\xi$ is in $R_{TF}$ units)
\begin{equation}
\xi^2 = 1/(2 m^* C(T) |\Delta|^2) ,\label{healwrong}
\end{equation}
where $|\Delta|^2$ is the ``density'' of Cooper-pairs
(in $k_B T_c$ units). In the case of the TF approximation for 
BEC, the order parameter is
substituded by its value in the middle of the trap.
Making a corresponding substitution here
gives $\xi^2 =\frac{1}{2m^*E_\mu(T)}$ which is the same as the
definition of correlation length in metallic superconductors in the
GL regime. However, it does not correctly describe the numerically 
obtained values for the healing length because, for the experimentally
feasible parameters used here, the energy scales in the GL equation
do not correspond to the TF limit of BEC. 

We propose a measure for the vortex core size by demanding that the 
kinetic plus potential energy term in Eq.(\ref{2dim}) has its minimum
value. This term corresponds to a vortex in a harmonic trap, that is,
the first excited state. The energetically favoured position is thus
the maximum of the first excited state wavefunction which coincides
with the oscillator ground state length. We thus define
\be
\xi^2=\frac{1}{m^* \omega^*} = \sqrt{\frac{7\zeta(3)}{24\pi^2}}\left(\frac{T_c}{T}\right)
\left(\frac{\Omega}{k_B T_c}\right)\sqrt{\frac{\lambda}{2\lambda+1}} .
\label{heal}
\ee
Figure 1 shows the order parameter for a selected set of
parameters, together with the two estimates Eq.(\ref{healwrong}) and 
Eq.(\ref{heal}) for the healing length. The results fit excellently
with Eq.(\ref{heal}) whereas the deviation from Eq.(\ref{healwrong})
is considerable and qualitatively different for small and large $\Delta$.
The definition of Eq.(\ref{heal})
is a function of temperature decreasing as T approaches $T_c$
(we have confirmed this temperature dependence also numerically). 
This is an opposite behaviour to metallic superconductors where 
$\xi \sim 1/\sqrt{1-T/T_c}$. 
The trapping energy becomes relatively 
stronger as $T\rightarrow T_c$ because $\Delta$ decreases.
This means stronger confinement for $\Delta(\rho)$ and 
decreasing $\xi$. Note that also in BEC, the confinement 
determines the healing length if one is away from the 
TF regime \cite{Fetter2}. Differences to BEC arise, however, from
the temperature dependence of the GL equation and from the
normalization.
     
{\it Critical temperature for a trapped system} 
As pointed out in \cite{Baranov} one can use 
the GL equation to estimate
the critical temperature of the trapped
system, as compared to the corresponding homogeneous case. Close to
$T_c$ the non-linear term is negligible and the GL equation reduces
to the Schr\"odinger equation for a trapped particle, $E_\mu(T)$ now
denoting the energy. The smallest possible energy $E_\mu(T_c) = \frac{3}{2}
\tilde{\omega}$ is then simply the ground state energy of the trap,
$\omega^* +\frac{1}{2}\omega_z^*$. Equating $\frac{3}{2}\tilde{\omega} = \omega^* +
\frac{1}{2}\omega_z^*$ gives $T_c$ in terms of $T_c^{(0)}$, $\omega^*$
and $\omega^*_z$ by the definition 
$\tilde{\omega} =\frac{2}{3} \ln (T_c^{(0)}/T_c)$. We use this to calculate
the new critical temperatures when the effects of rotation and 
Hartree-fields are added. Below we estimate the temperature regimes
for vortex solutions using a similar argument.

{\it Temperature regimes for the existence of a vortex solution}
The ground state energy of the trapped superfluid in the quasi-linear
regime can be estimated as above to be $E_\mu(T_c) = \omega^*$ (in
2-D). A vortex has a higher energy, and $E_\mu(T) = \ln (T_c/T) + \omega^*$
must have large enough value in order the GL equation to have a solution.
The minimum extra energy that
a vortex in a trapped non-interacting superfluid requires is the 2-D 
harmonic oscillator energy $\omega^*$
(this is also the vortex energy for a non-interacting 
BEC \cite{Dalfovo-Stringari}). Thus the maximum temperature $T_v$ at which
$E_\mu(T)$ can provide this extra energy is given by $\ln (T_c/T_v) =
\omega^*$. We have checked the validity of this estimate 
for a system of $N=3*10^5$ atoms,
scattering lentgh $|a|=1140$\AA, and trap frequency
$\Omega=820Hz$. 
The maximum T at which a vortex solution can exist, 
estimated by $T_c=T_ve^{\omega^*}=T_ve^{0.0146\frac{T_c}{T_v}}$, 
is $T_v \sim 0.98T_c$. 
The actual maximum T where the GL equation gives a numerical solution
was about $0.97T_c$. The deviation is due to neglecting the non-linear
term, and $T_v$ can be understood as the upper bound for the maximum
temperature.

{\it Hartree-fields and rotation}
In practical systems, the Cooper-paired atoms are always interacting
with the normal part of the gas -- whose density distribution is now 
position dependent. Moreover, the whole
system may be rotating, c.f.\ vortices in atomic BEC \cite{boson}. 
The GL equation used in this paper was
derived in \cite{Baranov} using the TF
approximation \cite{Tf} for
the density profile of the trapped Fermi-gas in a harmonic symmetric 
potential. We follow this derivation and add a new potential term 
$V({\bf r})$ in the local density approximation. This potential can
describe e.g.\ the Hartree fields or rotation. It is assumed to be
small enough in the sense that one can
still use the quasi-classical expression for the product of two Green's
functions in Eq.(7) of \cite{Baranov}.
We obtain the GL solution as given by Eq.(9) of \cite{Baranov} but with a new
term in the expansion of the Fermi energy around R=0 (see also \cite{StoofFP})
 given by
$\varepsilon_F(R)/\varepsilon_F= 1-R^2-V(R)$,
where $R={\bf r}/R_{TF}$ and the $R_{TF}=v_F/\Omega$.
Note that $R_{TF}$ and $\Omega$ might have changed because 
of adding $V({\bf r})$.
Expanding to second order and derivating with respect to 
$\Delta^{\ast}$ the ``potential term'' in the GL equation now reads
\begin{equation}
\frac{1+2\lambda}{2\lambda}(R^2+V^{(2)}(R)) , \label{v2}
\end{equation}
where $V^{(2)}(R)$ denotes an expansion of $V$ to second order.
The potential $V(R)$ has to be smooth enough for a second order expansion to
be sufficient. 

{\it Effect of rotation on the critical temperature and the healing length}
The TF approximation with a rotation term 
$V(\rho)= - \omega L_z = - \frac{1}{2}m\omega^2\rho^2$
gives 
$\varepsilon_F(R)/\varepsilon_F=
1-{\rho}^2-\left({z\gamma}\right)^2$
in the rotating frame of reference, where now
$R_{TF}= v_F/\sqrt{(\Omega^2-\omega^2)}$ and
$\gamma=(A\Omega^2/(\Omega^2-\omega^2))^{1/2}$,
and A is the assymetry in the trapping potential as defined earlier.
This means that both 
$m^*$ and $\omega^*$ change.
As the trapping frequency of the 
atoms is smaller now, $\Omega_{ho}^r=((\Omega^2-\omega^2)\Omega)^{1/3}$,
the Fermi energy and
$\lambda$ decrease: we define $\lambda^r=\lambda[1-\left(\frac{\omega}
{\Omega}\right)^2]^{1/6}$. 
This gives the new homogeneous system
critical temperature 
$T_c^{r(0)}$, and
using the same kind of procedure as above for a non-rotating system,
the new healing length $\xi_r$ and critical temperature $T_c^r$
can be calcutated. In the limit 
$\left(\frac{\omega}{\Omega}\right)^2 \rightarrow 1$, 
the critical temperature $T_c^{r(0)}\rightarrow 0$ as for BEC \cite{Stringari}, and  
$\xi^r$ as well as $\ln({T_c^{r(0)}}/{T_c^r})$ diverge.

{\it Effect of the normal fermions}
The TF approximation including the 
non-paired fermions was
introduced in \cite{Stoofnor} giving 
$\frac{\varepsilon_F(R)}{\varepsilon_F}= 1-R^2+\frac{1}{\varepsilon_F}
\frac{4\pi\hbar^2 |a|}{m}n(R)$, 
where $n(R)$ is the density distribution of the atoms in one hyperfine
state. It was shown that the
Hartree-field increases the critical temperature of the system. 
This result is also given by our GL treatment:
Consider again the weakly non-linear regime close to $T_c$.
We have a Schr\"odinger equation for a spherically 
symmetric potential but with a new term given by the Hartree field.
Because of the smooth shape of the 
fermionic distribution one can assume $n(R) \sim n^{(2)}(R)$ and 
use $V^{(2)}(R) \simeq -\frac{1}{\varepsilon_F}
\frac{4\pi\hbar^2 |a|}{m}n(R)$ in Eq.(\ref{v2}). 
Thus we have effectivelly 
a new harmonic symmetric potential $m^*(\omega^*)^2R^2/2-C'n(R) \sim 
-E_{HF}+m^{*}(\omega^{*HF})^2R^2/2$ with 
$s=\left(\omega^{*HF}/\omega^*\right)^2 > 1$. That is,
 the potential is now deeper and with higher 
``frequency''. Because of the new higher Fermi energy
$\varepsilon_F^{HF}=\varepsilon_F+E_{HF}$ 
the critical temperature in the corresponding homogeneous system,
$T_c^{(0)HF}$, increases. 
Simple considerations give
$\ln (T_c^{HF(0)}/T_c^{HF})=\ln (T_c^{(0)}/T_c)s^{1/2}$.
This means that $T_c^{HF}$ actually deviates more from $T_c^{HF(0)}$ than
$T_c$ from $T_c^{(0)}$; the effect of trapping is enhanced because also
the normal fermions feel the trapping potential. On the other hand, 
$T_c^{HF(0)} > T_c^{(0)}$, and for the parameter values we have considered
the total effect is that the Hartree fields increase the critical
temperature ($T_c^{HF} \sim 2 T_c$).
 
{\it Self-consistent solution of normal and superfluid fermions}
As a non-rigorous but intuitive first guess towards a self-consistent
treatment, we calculate $n(r)$, instead of the TF approximation, by 
the BCS theory in local density approximation but with $\Delta (r)$
given by the GL equation:
\be
n(r) = \int \frac{d^3k}{(2\pi)^3}\left[ |u_k(r)|^2f(E_k) +
|v_k(r)|^2\left(1 - f(E_k)\right)\right] ,
\ee
where $|u_k(r)|^2,|v_k(r)|^2=1/2[1 \pm \xi_k/\sqrt{\xi_k^2+|\Delta(r)|^2}]$ and
$\xi_k(r)= \frac{\hbar^2k^2}{2m}-\frac{4\pi\hbar^2|a|}{m}n(r)
-(\mu-\frac{1}{2}m\Omega^2r^2)$,
and $f$ is the Fermi-Dirac distribution.
We solve $n(r)$ from this equation, use it in the GL
equation, and iterate  
until sufficient convergence is found.

The results of the self-consistent calculation are presented
in Fig.2. The order parameter increases considerably, mainly because
of the increase in the critical temperature when the Hartree field is
added ($T$ was fixed). Quasiparticles fill  
the vortex core: there is about $5\%$ increase of $n(0)$ compared to $n(\xi)$.
It would be interesting to compare our simple self-consistent
treatment to the rigorous description by the BdG equations.

{\it Observation of a vortex} For the parameters used in our calculations,
the vortex core sizes/healing lengths vary between 2-10 $\mu$m. This
is close to but still above the diffraction limit of light.
In principle, for instance the laser probing method of \cite{ours}
could be extended to the observation of a vortex: the applied Raman
beams are focused so that they intersect either only in the core, or only in the
superfluid region, and these two choices give absorption peaks at different
frequencies.  

Also the smallness of the superfluid fraction makes the observation a
challenge. In our results, the maximum value of the order parameter
(vortex height) is of the same order of magnitude as the temperature. 
To estimate how it depends on temperature
and other parameters, we approximate $\Delta_{max} \sim \Delta (\rho
\sim \xi)$: 
we insert $\rho = \xi$ into the GL equation, 
neglect the first derivative (maximum) and approximate the second derivative 
by assuming a parabolic shape of $\Delta (\rho)$.
This gives (for comparison with the numerics, see Fig.1)
\begin{equation}
\Delta_{max} \sim \sqrt{(\ln (T_c/T) - \omega^*)/C(T)} \label{dm}. 
\end{equation}
Note that our estimate for the maximum vortex temperature $T_v$ corresponds
to $\Delta_{max}$ being real.

We have used the Ginzburg-Landau equation in a trapped geometry to define
analytical estimates for the basic quantities describing a trapped
superfluid Fermi-gas: the vortex core size and height, maximum temperature for a
vortex solution, and the changes caused by additional potentials such as
rotation or spatially varying Hartree fields.  
A striking difference to metallic superconductors was found in the
temperature and system parameter dependence of the
vortex core size/healing length.  
Our results indicate that the effect of the confining geometry is
essential for mesoscopic fermionic superfluids, especially when
considering excited state solutions such as vortices.

{\it Acknowledgements} We thank the Academy of Finland for support
(projects 42588, 48845, 47140 and 44897). G-S.P.\ also acknowledges
NSF support through the project NSF DMR 99-86199l.

\narrowtext
\vbox{
\begin{figure}
\begin{center}
\epsfig{file=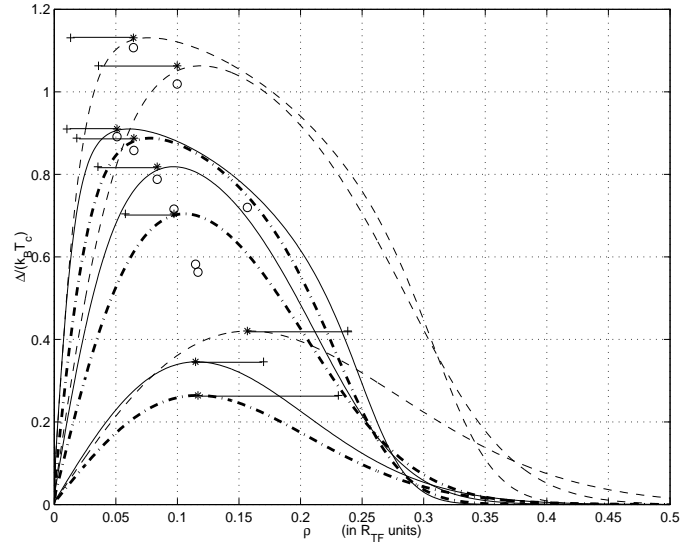,height=0.31\textheight,width=0.5\textwidth}
\end{center}
\caption{Vortex solutions for the order parameter in $k_BT_c$ units and
in trap units. Solid lines are for
$\Omega=2\pi100Hz, 1kHz, 3kHz$ ($N=3*10^5$, $|a|=1140$\AA, $T=0.89T_c$).
Dashed lines correspond to $N =10^5, 3*10^5, 10^6$ ($T=0.8T_c$,
$\Omega=820Hz$ and $|a|=1140$\AA). Dot-dashed lines are for 
$|a| = 985, 1118, 1608$\AA\ ($N=3*10^5$, $\Omega=820 Hz$,
$T=0.89T_c$). In all the three cases, the highest value of $\Omega$,
$N$ or $|a|$ corresponds to the curve with largest maximum $\Delta$.
The healing lengths given by Eq.(\ref{heal}) and 
Eq.(\ref{healwrong}) are represented by * and +, respectively. The estimate
for $\Delta_{max}$ given by Eq.(\ref{dm}) is represented by $o$.}
\label{casos}
\end{figure}
}
\narrowtext
\vbox{
\begin{figure}
\begin{center}
\epsfig{file=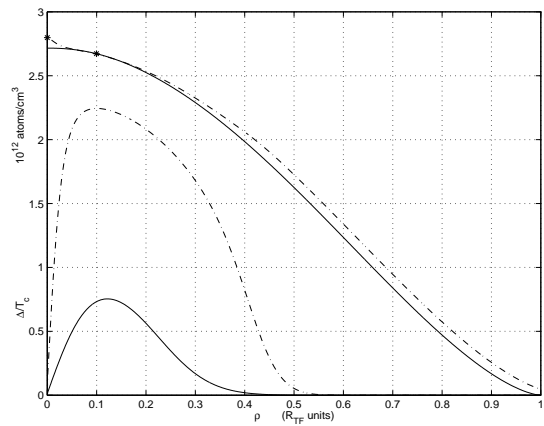,height=0.24\textheight,width=0.4\textwidth}
\end{center}
\caption{Vortex solutions for the order parameter are given in $k_B T_c$ units 
and density of normal fermions in one hyperfine state in $10^{12} 
atoms/cm^3$. Solid lines are the independent initial solutions for Eq.(11) 
and the GL
equation, the dot-dashed lines are the self-consistent solutions.
Here $T=0.9T_c$, $N \sim 3*10^5$, $\Omega=820Hz$ and $|a|=1140$\AA.}
\label{hf}
\end{figure}
}
\end{multicols}

\begin{thebibliography}{cc}
\bibitem{Debbie}
B.\ DeMarco and D.S.\ Jin, Science \textbf{285}, 1703 (1999); 
M.J.\ Holland {\it et al.}, Phys.\ Rev.\ A \textbf{61}, 053610 (2000).
\bibitem{Salomon}
M.O.\ Mewes {\it et al.}, Phys.\ 
Rev.\ A \textbf{61}, 011403 (R) (2000); see also cond-mat/0011291.
\bibitem{Thomas}
K.M.\ O'Hara {\it et al.}, Phys.\ Rev.\ Lett.\ \textbf{85}, 2092 (2000). 
\bibitem{Stoof}
H.T.C.\ Stoof {\it et al.} , 
Phys.\ Rev.\ Lett.\ \textbf{76}, 10 (1996).  
\bibitem{Holland}
M.\ Holland {\it et al.}, cond-mat/0103479.
\bibitem{bcsbec}
See M.\ Randeria and references therein in {\it Bose-Einstein Condensation},
eds.\ A.\ Griffin, D.W.\ Snoke, and S.\ Stringari (Cambridge Un.\ Press,
Cambridge, 1995). 
\bibitem{all}
W.\ Zhang {\it et al.} Phys.\ Rev.\ A \textbf{60}, 504 (1999); 
J.\ Ruostekoski, Phys.\ Rev.\ A \textbf{60}, R1775 (1999); 
F.\ Weig and W.\ Zwerger, Europhys.\ Lett.\ \textbf{49}, 282 (2000); M.A.\ Baranov
 and D.S.\ Petrov, 
Phys.\ Rev.\ A \textbf{62}, 041601(R) (2000); M.\ Farine {\it et al.},
 Phys.\ Rev.\ A \textbf{62}, 013608 (2000); G.M.\ Bruun and C.W.\
 Clark, J.\ Phys.\ B \textbf{33}, 3953 (2000).
\bibitem{ours}
P.\ T\"orm\"a and P.\ Zoller, Phys.\ Rev.\ 
Lett.\ \textbf{85}, 487 (2000) and G.M.\ Bruun {\it et al.},
cond-mat/0011333. 
\bibitem{He}
R.J.\ Donelly, {\it Quantized vortices in helium II}, (Cambridge
University Press, 1991).
\bibitem{boson}
For a review see e.\ g.\ A.L.\ Fetter and A.A.\ Svidzinsky, 
J.\ of Phys.: Cond.\ Mat.\ {\bf 13}, R135 (2001)
and references therein. 
\bibitem{BdG}
P.\ de Gennes, {\it Superconductivity of metals and alloys} (Addison- Wesley,
New York, 1966).
\bibitem{Ivan}
G.M.\ Bruun {\it et al.}, Europhys.\ D \textbf{7}, 433
(1999); A.\ Minguzzi {\it et al.}, cond-mat/0103591.
\bibitem{Baranov}
M.A.\ Baranov and D.S.\ Petrov, Phys.\ Rev.\ A \textbf{58}, R801 (1998). 
\bibitem{Tf}
D.A.\ Butts and D.S.\ Rokshar, Phys.\ Rev.\ A \textbf{55}, 4346 (1997).
\bibitem{method}
F.\ Dalfovo and M.\ Modugno, Phys.\ Rev.\ A \textbf{61}, 023605 (2000). 
\bibitem{Pethick}
G.\ Baym and C.J.\ Pethick, Phys.\ Rev.\ Lett.\ \textbf{76}, 6 (1996).
\bibitem{Fetter2}
A.L.\ Fetter, cond-mat/9811366, chapter II.C.
\bibitem{sca}
E.R.I.\ Abraham {\it et al.}, Phys.\ Rev.\ A \textbf{55}, R3299 (1997). 
\bibitem{Dalfovo-Stringari}
F.\ Dalfovo and S.\ Stringari, Phys.\ Rev.\ A \textbf{53}, 2477 (1996).
\bibitem{StoofFP}
M.\ Houbiers and H.T.C.\ Stoof, Phys.\ Rev.\ A \textbf{59}, 1556 (1999).
\bibitem{Stringari}
S.\ Stringari, Phys.\ Rev.\ Lett.\ \textbf{82}, 4371 (1999).
\bibitem{Stoofnor}
M.\ Houbiers {\it et al.}, Phys.\ Rev.\ A \textbf{56}, 4864, (1997).
\end{thebibliography}
\end{document}